\documentclass[varenna]{cimento}

\title{Latest results on gamma-ray pulsars with {\slshape Fermi}}

\author{P.~M.~Saz Parkinson}

\institute{Santa Cruz Institute for Particle Physics, University of California, Santa Cruz, CA 95064, USA}

\institute{Department of Physics, Laboratory for Space Research, University of Hong Kong, Hong Kong}

\shortauthor{P.~M.~Saz Parkinson}

\begin{document}

\maketitle

\section{Gamma-ray pulsars pre-{\slshape Fermi}}

Gamma-ray astronomy has a long history, going back to the 1960s. In
the 1990s, the Energetic Gamma Ray Experiment Telescope (EGRET), on
board the {\it Compton Gamma-Ray Observatory} (CGRO\footnote{The second of
    NASA's great observatories.}, 1991--2000), detected almost 300
gamma-ray sources, a majority of which were
unidentified~\cite{3EG}. The seven gamma-ray pulsars detected by CGRO (6 by
EGRET, and PSR~B1509--58 by COMPTEL), shared many characteristics
(e.g. young and highly energetic, mostly double-peaked) but also covered various categories:
{\it radio-loud}, {\it radio-quiet} (Geminga), {\it soft} (MeV) , but the somewhat 
limited statistics (particularly above 5 GeV), made it
challenging to discriminate between the leading pulsar emission models~\cite{Thompson04}. For a 
review of the EGRET era results, immediately preceding the launch of
{\it Fermi}, see Thompson (2008).

\section{The {\slshape Fermi} era}

The {\it Fermi} gamma-ray space telescope, launched on 11 June 2008,
is a giant leap forward for gamma-ray astronomy. The Large Area
Telescope~\cite{Atwood09} (LAT), the main instrument on {\it Fermi}, uses silicon
strip detectors (far superior to the old gaseous detectors), making it
the most sensitive instrument in the $\sim$0.5--300 GeV energy range for the foreseeable
future. Indeed, the LAT recently detected its billionth gamma ray ($\sim$1000 times the
number of gamma rays detected by EGRET in its lifetime) and is showing
no signs of aging. Not surprisingly, {\it Fermi} quickly made a big
impact in many areas, and pulsars in particular, for example
uncovering a large population of {\it radio-quiet} gamma-ray pulsars
hiding among the {\it previously unidentified} EGRET
sources~\cite{BSP1,BSP2}. The most recent catalog released by the LAT Collaboration, the Third LAT
source catalog (3FGL), contains over 3,000 sources, of which approximately one
third are {\it unassociated}~\cite{3FGL}. Uncovering the nature of
LAT {\it unassociated} sources is (and will remain for many years) a
key pursuit for the gamma-ray (and broader) astrophysics community. In
this context, a number of statistical methods (e.g. machine learning techniques,
neural networks), in combination with multi-wavelength follow-up
observations are helping to identify the likely nature of many of
these sources~\cite{Chiaro16,Saz16}. For a detailed review of the
``Gamma-ray Pulsar Revolution'', see \cite{Caraveo14}.

\subsection{Recent pulsar results with Pass 8}

The event selection algorithms developed for the LAT are the result of a long,
iterative process, with the various releases known as {\it
  Passes}. {\it Pass 6} data were publicly released after launch but based only on {\it pre-launch}
information. {\it Pass 7} data, released in August 2011, incorporated
knowledge gained from the first few years in orbit. The {\it Pass 8}
release represents a complete redesign of every aspect of the event
selection, leading to a significant increase in effective
area, an improvement in the point-spread function, and a reduction in
background contamination~\cite{Pass8}. Because every {\it Pass}
results in the entire {\it Fermi} data (from the beginning of the
mission) being reprocessed, the release of {\it Pass 8} produced
scientific results immediately after its release, without the need to
wait for {\it additional} data. 

A significant number of known pulsars suddenly showed gamma-ray pulsations with {\it Pass
  8}, despite being previously undetected~\cite{Laffon15}. The {\it Pass
  8} data also improved significantly the sensitivity of LAT blind
searches for pulsars. The Einstein@Home survey, for example,
recently reported 17 new (mostly radio-quiet) gamma-ray
pulsars~\cite{Einstein}. The number of gamma-ray pulsars detected by
{\it Fermi} (now over 200) continues to increase, with the rate of
discovery showing no signs of tapering off\footnote{{\small{\tt
      https://confluence.slac.stanford.edu/x/5Jl6Bg}}}. Interestingly,
millisecond pulsars (MSPs) represent roughly half the entire gamma-ray
pulsar population, with some of them meeting the stringent criteria to be added to
the pulsar timing arrays, thus aiding in the search for graviational
waves~\cite{Ray12}. One of the most interesting new gamma-ray pulsars detected by the
LAT is PSR J0540--6919, in the Large Magellanic Cloud (LMC), located at
$\sim$50 kpc, making it the first extra-Galactic gamma-ray pulsar (and hence
the most distant) ever detected~\cite{LMCPulsar}. Curiously, PSR
J0537--6910, also in the LMC and with very similar characteristics,
still shows no gamma-ray pulsations. 

\subsection{Gamma-ray binaries with Fermi}

Another gamma-ray source in the LMC that has recently attracted a
great deal of attention was first identified, rather mundanely, as
P3~\cite{LMC}. This source turns out to be a gamma-ray
binary with a 10.3 day orbital period, as confirmed also by radio and X-ray observations~\cite{Corbet16}. Coming over
four years after the discovery of 1FGL J1018.6--5856 (J1018), the first gamma-ray binary
discovered by {\it Fermi}~\cite{Corbet11, J1018}, this new gamma-ray
binary broke several records (most luminous gamma-ray binary, first
extra-Galactic gamma-ray binary), and like J1018, is likely powered by an
energetic pulsar~\cite{Corbet16}. 

While many (if not most) gamma-ray binaries are thought to contain
pulsars, in most cases the pulsar has eluded detection
(e.g. LSI+61$^\circ$303, LS 5039). In one recent case, however, the pulsar
(J2032+4127) was discovered {\it first}, while the binary nature of the
system was uncovered subsequently. When first discovered in a blind search by {\it
  Fermi}, PSR J2032+4127 was thought to be an isolated gamma-ray pulsar~\cite{BSP1}. Long-term timing in radio,
however, reveals it to be in a binary system with a very long
($\sim$decades) orbital period~\cite{Lyne15}. Recent multi-wavelength monitoring
observations report an increase in X-ray emission from the
system (by a factor of $\sim$20 since 2010 and a factor of $\sim$70
since 2002) and refined its orbital period to be 45--50 yr, with
its time of periastron predicted to be in November 2017~\cite{Ho17}.

The LAT has also been very successful at finding so-called ``black widow'' or
``redback'' systems: eclipsing binary millisecond pulsars {\it eating} away their low-mass companion star, with their
radiation beams. Some of these systems are first identified through their multi-wavelength
emission, such as the case of 0FGL
J2339.8--0530~\cite{Romani11}. Radio follow-up searches in this case
revealed a pulsar~\cite{Ray14} and gamma-ray pulsations were also detected~\footnote{See talk by A. Belfiore at the 2013 Aspen Meeting on
Physical Applications of Millisecond Pulsars, {\tt
  http://aspen13.phys.wvu.edu/aspen\_talks/Belfiore\_Gamma\_Ray\_Searches.pdf}}. Long
term gamma-ray timing of PSR J2339--0533 recently revealed dramatic orbital-period modulations ascribed
to a change in the gravitational quadrupole moment~\cite{Pletsch15}. Due to the eclipsing nature of these systems,
radio non-detections are frequent, making gamma-ray searches
complementary. Indeed, in one case, the pulsar was discovered in gamma
rays first~\cite{Pletsch12}, with radio pulsations coming later~\cite{Ray13}.  A number of redback candidates have been
identified (e.g. 3FGL J2039.6--5618~\cite{Romani15,Salvetti15}, 3FGL
J0212.1+5320~\cite{Li16,Linares17})  and searches for these pulsars are ongoing.

\subsection{Variable and transition gamma-ray pulsars}

Until recently, gamma-ray pulsars were thought to be {\it steady}
sources\footnote{In fact, a key characteristic distinguishing pulsars from AGN is precisely the {\it low
    variability}}. The long-term monitoring of large numberes of
pulsars over a period of years, however, has started to reveal more
complicated behavior in some sources. PSR J2021+4026, a bright, {\it
  radio-quiet} gamma-ray pulsar discovered by {\it Fermi} early on in
the mission~\cite{BSP1} experienced an abrupt drop in flux of
$\sim$20\%, associated with a $\sim$4\% increase in spindown rate,
also accompanied by changes in the pulse profile, making this the
first known variable gamma-ray pulsar~\cite{Allafort13}. The most
recent observations appear to show that the flux of J2021+4026 has now
gone back to its original values~\cite{Ng16}. 

An even more dramatic transition was detected in PSR J1023+0038, the so-called ``missing
link'' pulsar known to have previously been in a Low Mass X-ray Binary
state, subsequently switching to a rotation-powered state. Recently,
this system experienced new state transition, with a five-fold
increase in gamma-ray flux accompanying the disappearance of the radio
pulsations~\cite{Stappers14}.

Another pulsar that has benefitted from the long-term monitoring
capabilities of the LAT is PSR J1119--6127~\cite{Camilo00}.  This young, energetic
pulsar associated with supernova remnant G292.2--0.5 has an extremely large inferred surface
magnentic field ($\sim4\times10^{13}$G), and was detected as a
gamma-ray pulsar early on by the LAT~\cite{Parent11}. Recently, the
{\it Fermi} GBM~\cite{GCN19736} and {\it Swift}~\cite{ATel9274} detected a series of strong SGR-like
bursts, followed by hard X-ray pulsations~\cite{ATel9282}, in conjunction with
a large spin-up glitch~\cite{ATel9284}. Radio pulsations disappeared~\cite{ATel9286}, reappearing two weeks
later~\cite{ATel9366}. Unfortunately, despite a one-week LAT Target of Opportunity
(TOO) pointed observation (increasing the exposure by a factor of
$\sim$2.4), no significant changes in gamma-ray flux were
detected~\cite{ATel9365}, and no significant pulsations were detected
post-burst~\cite{ATel9378}.

Finally, the recent possible detection of pulsed {\it soft} gamma-ray emission from PSR
J1846--0258 (up to 100 MeV) is of great interest~\cite {ATel9077}. This
pulsar shares many similarities with PSR J1119--6127: large magnetic
field and past {\it magnetar-like} bursts following a large
glitch. Thus, it represents another possible ``transition'' pulsar,
making it a worthwhile target to monitor, going forward.

\section{Conclusions}
Since its launch, almost nine years ago, {\it Fermi} has produced a
long list of discoveries in the field of gamma-ray pulsars. More
surprisingly, the rate of these discoveries does not appear to be
slowing down. {\it Fermi} continues to detect new pulsars in every category: young, MSPs, radio-loud,
radio-quiet, etc. Finally, the longer data sets and the development of {\it
  Pass 8} are now enabling {\it Fermi} to delve deeper into new
parameter space, revealing a range of {\it variability} in gamma-ray
pulsars that was hitherto unknown.

\acknowledgments

I thank the organizers of SciNeGHE 2016 for putting together such an interesting and timely conference. In particular, I
am very grateful to Max Razzano, for his hospitality during my visit to Pisa. The \textit{Fermi}-LAT Collaboration acknowledges support for LAT
development, operation and data analysis from NASA and DOE (United States), CEA/Irfu and IN2P3/CNRS (France), ASI and INFN (Italy), MEXT,
KEK, and JAXA (Japan), and the K.A.~Wallenberg Foundation, the Swedish
Research Council and the National Space Board (Sweden). Science analysis support in the operations phase from INAF (Italy) and CNES
(France) is also gratefully acknowledged. 
%This work performed in part under DOE Contract DE-AC02-76SF00515.

\end{document}